\documentclass[prl, aps, reprint, twocolumn, superscriptaddress, floatfix]{revtex4-1}

\newcommand{\ket}[1]{\left|#1\right\rangle}

\usepackage{cancel}
\usepackage{amsmath}
\usepackage[separate-uncertainty = true,multi-part-units=single]{siunitx}
\usepackage{bbold}
\usepackage{graphicx}
\usepackage{blkarray}
\usepackage{sidecap}

\usepackage{titlesec}
\usepackage{setspace}

\usepackage{soul}
\usepackage{color}

\titleformat{\section}{\bfseries}{}{0em}{}[]
\titlespacing{\section}
  {0pt}
  {2.0ex plus 1ex minus 1ex}
  {3pt}
\titleformat*{\subsection}{\normalsize\itshape}
\titlespacing{\subsection}
  {0pt}
  {2.0ex plus 1ex minus 1ex}
  {3pt}

\begin{document}

\title{\vspace{-5ex} \large \textbf{Searching for Dark Matter with a Superconducting Qubit}}

\author{Akash V. Dixit}
\email{avdixit@uchicago.edu}
\affiliation{James Franck Institute, University of Chicago, Chicago, Illinois 60637, USA}
\affiliation{Department of Physics, University of Chicago, Chicago, Illinois 60637, USA}
\affiliation{Kavli Institute for Cosmological Physics, University of Chicago, Chicago, Illinois 60637, USA}

\author{Srivatsan Chakram}
\affiliation{James Franck Institute, University of Chicago, Chicago, Illinois 60637, USA}
\affiliation{Department of Physics, University of Chicago, Chicago, Illinois 60637, USA}
\affiliation{Department of Physics and Astronomy, Rutgers University, Piscataway, New Jersey 08854, USA}

\author{Kevin He}
\affiliation{James Franck Institute, University of Chicago, Chicago, Illinois 60637, USA}
\affiliation{Department of Physics, University of Chicago, Chicago, Illinois 60637, USA}

\author{Ankur Agrawal}
\affiliation{James Franck Institute, University of Chicago, Chicago, Illinois 60637, USA}
\affiliation{Department of Physics, University of Chicago, Chicago, Illinois 60637, USA}
\affiliation{Kavli Institute for Cosmological Physics, University of Chicago, Chicago, Illinois 60637, USA}

\author{Ravi K. Naik}
\affiliation{Department of Physics, University of California Berkeley, California 94720, USA}

\author{David I. Schuster}
\affiliation{James Franck Institute, University of Chicago, Chicago, Illinois 60637, USA}
\affiliation{Department of Physics, University of Chicago, Chicago, Illinois 60637, USA}
\affiliation{Pritzker School of Molecular Engineering, University of Chicago, Chicago, Illinois 60637, USA}

\author{Aaron Chou}
\affiliation{Fermi National Accelerator Laboratory, Batavia, Illinois 60510, USA}

\begin{abstract}
  Detection mechanisms for low mass bosonic dark matter candidates, such the axion or hidden photon, leverage potential interactions with electromagnetic fields, whereby the dark matter (of unknown mass) on rare occasion converts into a single photon. Current dark matter searches operating at microwave frequencies use a resonant cavity to coherently accumulate the field sourced by the dark matter and a near standard quantum limited (SQL) linear amplifier to read out the cavity signal. To further increase sensitivity to the dark matter signal, sub-SQL detection techniques are required. Here we report the development of a novel microwave photon counting technique and a new exclusion limit on hidden photon dark matter. We operate a superconducting qubit to make repeated quantum non-demolition measurements of cavity photons and apply a hidden Markov model analysis to reduce the noise to $\SI{15.7} {\dB}$ below the quantum limit, with overall detector performance limited by a residual background of real photons. With the present device, we perform a hidden photon search and constrain the kinetic mixing angle to $\epsilon \leq 1.68 \times 10^{-15}$ in a band around $\SI{6.011} {\giga \hertz}$ ($ \SI{24.86} {\micro \electronvolt}$) with an integration time of $\SI{8.33} {\second}$. This demonstrated noise reduction technique enables future dark matter searches to be sped up by a factor of 1300. By coupling a qubit to an arbitrary quantum sensor, more general sub-SQL metrology is possible with the techniques presented in this work.

\end{abstract}

\maketitle

\section {Introduction}
The nature of dark matter is an enduring mystery of our universe. Observations of galaxy rotation curves, gravitational lensing, and the presence of structure in the cosmos all inform our understanding of dark matter, but provide little insight into its intrinsic properties \cite{Tanabashi_2018, Rubin_1980}. Though the gravitational evidence for the existence of dark matter is extensive \cite{Rubin_1980}, thus far, dark matter has evaded direct detection in terrestrial experiments. We are interested in testing the hypothesis that dark matter is composed of waves of low mass bosons, which due to their high galactic phase space density, arrive as coherent waves with macroscopic occupation number. Well known dark matter candidates include the axion and hidden sector photon, which both have compelling cosmological origin stories \cite{Preskill_1982, Abbott_1982, Dine_1982, Arias_2012, Graham_2016}.

One method for detecting these dark matter waves exploits their interactions with the electromagnetic field \cite{Sikivie_1983, Graham_2016}. A microwave cavity with resonance frequency tuned to the mass of the hypothetical particle is used to coherently accumulate the electromagnetic response (see Supplemental Material). On rare occasions, the dark matter deposits a single photon in the cavity.

There are specified targets in the parameter space of coupling and dark matter mass in the case of the axion of quantum chromodynamics (QCD). The expected signal photon occupation number is $\sim$10$^{-2}$ for searches like the Axion Dark Matter eXperiment operating at 650 MHz \cite{Du_2018}. However, for searches at higher frequencies, the microwave cavity volume must shrink to maintain the resonance condition. The signal photon rate scales with the volume of the cavity, making detection of smaller signals increasingly challenging at higher frequencies. For an axion search with the microwave cavity ($\SI{6.011} {\giga \hertz}$) used in the present work and given the experimental parameters in typical axion search experiments \cite{Braine_2020, Brubaker_2017, Zhong_2018, Backes_2021}, QCD axion models \cite{Dine_1981, Zhitnitsky_1980, Kim_1979, Shifman_1980} predict a signal with mean photon number of $\bar{n}_{\mathrm{axion}} \sim$10$^{-8}-10^{-5}$ per measurement. For hidden photons, the parameter space is less constrained, \cite{Arias_2012, Horns_2013, Chaudhuri_2015} and the mean photon number per measurement could be $\bar{n}_{\mathrm{HP}} \leq 10^{-1}$. Currently, these searches employ linear amplification operating near the standard quantum limit (SQL) \cite{Caves_1982} to read out the built up signal in the microwave cavity, where the noise variance is equivalent to fluctuations of an effective background of $\bar{n}_{\mathrm{SQL}} = 1$. At GHz frequencies and above, the noise inherent to quantum limited linear amplification overwhelms the signal, making the search untenable ($\bar{n}_{\mathrm{SQL}} \gg \bar{n}_{\mathrm{axion}}, \bar{n}_{\mathrm{HP}}$).

We use single photon resolving detectors to avoid quantum noise by measuring only field amplitude, resulting in insensitivity to the conjugate phase observable. The noise is then dominated by the Poisson fluctuations of the background counts and ultimately limited by the shot noise of the signal itself \cite{Lamoreaux_2013}. Superconducting nanowire single-photon detectors or photomultipier tubes can readily count infrared photons; however, these technologies are not well suited to detect single low energy microwave photons \cite{Hadfield_2009}. Here, we develop a detector that is sensitive in the microwave regime and has a low dark count probability commensurate with the small signal rates expected in a dark matter experiment.

\section {Qubit based photon counter}
\begin{figure}[t!]
    \centerline{
    \includegraphics[width=\columnwidth]{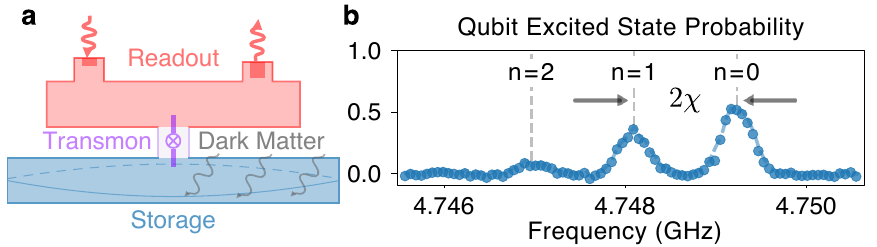}
    }
    \caption{\textbf{Superconducting transmon qubit dispersively coupled to high Q storage cavity}. \textbf{a,} Schematic of photon counting device consisting of storage and readout cavities bridged by a transmon qubit \cite{Leek_2010}. The interaction between the dark matter and electromagnetic field results in a photon being deposited in the storage cavity. \textbf{b,} Qubit spectroscopy reveals that the storage cavity population is imprinted as a shift of the qubit transition frequency. The photon number dependent shift is $2 \chi$ per photon.}
    \label{fig:device}
\end{figure}

In order to construct a single photon counter, we employ quantum non-demolition (QND) techniques pioneered in atomic physics \cite{Haroche_1990, Gleyzes_2007}. To count photons, we utilize the interaction between a superconducting transmon qubit \cite{Koch_2007, Ambegaokar_1963} and the field in a microwave cavity, as described by the Jaynes-Cummings Hamiltonian \cite{Jaynes_1963} in the dispersive limit (qubit-cavity coupling $\ll$ qubit, cavity detuning): $\mathcal{H}/\hbar = \omega_c a^{\dagger} a + \frac{1}{2}\omega_q \sigma_z + 2 \chi a^{\dagger} a \frac{1}{2}\sigma_z$. The Hamiltonian can be recast to elucidate a key feature: a photon number dependent frequency shift ($2 \chi$) of the qubit transition (Fig. \ref{fig:device}(b)).

\begin{equation}
\mathcal{H}/\hbar = \omega_c a^{\dagger} a + \frac{1}{2}(\omega_q + 2 \chi a^{\dagger} a) \sigma_z
  \label{eqn:hamiltonian}
\end{equation}

We use an interferometric Ramsey measurement of the qubit frequency to infer the cavity state \cite{Kono_2018}. Errors in the measurement occur due to qubit decay, dephasing, heating, cavity decay, and readout infidelity, introducing inefficiencies or worse, false positive detections. For contemporary transmon qubits, these errors occur with much greater probability (1-10\%) than the appearance of a dark matter induced photon, resulting in a measurement that is limited by detector errors. The qubit-cavity interaction ($2 \chi a^{\dagger} a \frac{1}{2}\sigma_z$) is composed solely of number operators and commutes with the bare Hamiltonian of the cavity ($\omega_c a^{\dagger} a$) and qubit ($\frac{1}{2}\omega_q \sigma_z$). Thus, the cavity state collapses to a Fock state ($\ket{0}$ or $\ket{1}$ in the $\bar{n}\ll1$ limit) upon measurement, rather than being absorbed and destroyed \cite{Braginsky_1996, Nogues_1999, Johnson_2010, Sun_2014}. Repeated measurements of the cavity photon number made via this QND operator enable us to devise a counting protocol, shown in Fig. \ref{fig:pulse_meas}(a), insensitive to errors in any individual measurement \cite{Zheng_2016, Hann_2018, Elder_2020}. This provides exponential rejection of false positives with only a linear cost in measurement time.

In this work, we use a device composed of a high quality factor ($Q_s = 2.06 \times 10^7$) 3D cavity \cite{Chakram_2020, Lei_2020} used to accumulate and store the signal induced by the dark matter (storage, $\omega_s = 2 \pi \times \SI{6.011} {\giga \hertz}$), a superconducting transmon qubit ($\omega_q = 2 \pi \times \SI{4.749} {\giga \hertz}$), and a 3D cavity strongly coupled to a transmission line ($Q_r = 1.5 \times 10^4$) used to quickly read out the state of qubit (readout, $\omega_r = 2 \pi \times \SI{8.052} {\giga \hertz}$) (Fig. \ref{fig:device}(a)). We  mount the device to the base stage of a dilution refrigerator at $\SI{8}{\milli \kelvin}$.

To count photons, we repeatedly map the cavity population onto the qubit state by performing a cavity number parity measurement with Ramsey interferometry, as depicted in Fig. \ref{fig:pulse_meas}(a). We place the qubit, initialized either in $\ket{g}$ or $\ket{e}$, in a superposition state $\frac {1}{\sqrt{2}} (\ket{g} \pm \ket{e}) $ with a $\pi/2$ pulse. The qubit state precesses at a rate of $|2 \chi| = 2 \pi \times \SI{1.13} {\mega \hertz}$ when there is one photon in the storage cavity due to the photon dependent qubit frequency shift. Waiting for a time $t_p = \pi/|2 \chi|$ results in the qubit state accumulating a $\pi$ phase if there is one photon in the cavity. We project the qubit back onto the z-axis with a $-\pi/2$ pulse completing the mapping of the storage cavity photon number onto the qubit state. We then determine the qubit state using its standard dispersive coupling to the readout resonator. For weak cavity displacements $(\bar{n} \ll 1)$, this protocol functions as a qubit $\pi$ pulse conditioned on the presence of a single cavity photon \cite{Kono_2018}. If there are zero photons in the cavity, the qubit remains in its initial state. If there is one photon in the cavity, the qubit state is flipped ($\ket{g} \leftrightarrow \ket{e}$). More generally, this protocol is sensitive to any cavity state with odd photon number population.

\section {Hidden Markov model analysis}
\begin{figure}[t!]
    \centerline{
    \includegraphics[width=\columnwidth]{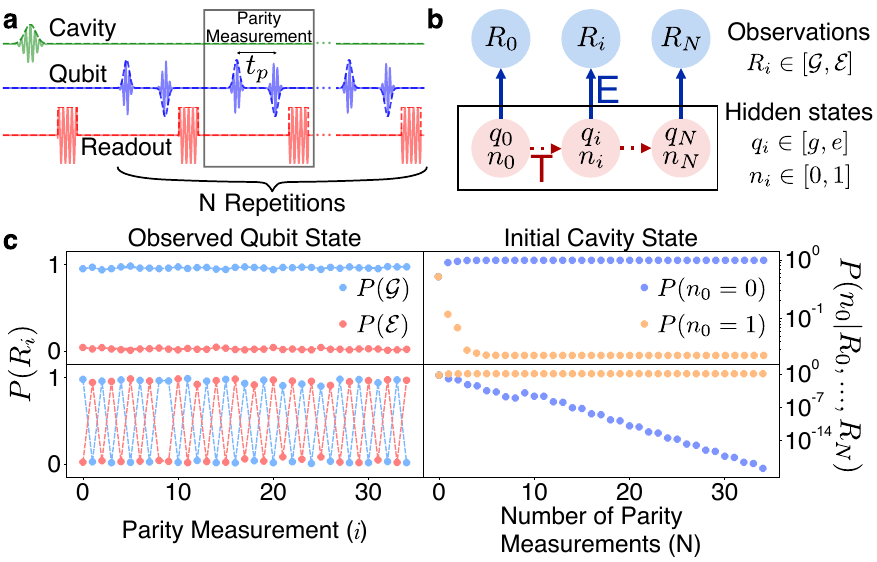}
    }
    \caption{\textbf{Photon counting protocol and hidden Markov model analysis}. \textbf{a,} Pulse sequence for photon counting includes cavity initialization and repeated parity measurements, consisting of a $\pi/2$ pulse, a wait time of $t_p$, and a $-\pi/2$ pulse followed by a qubit readout. \textbf{b,} Cavity and qubit states evolve under transition matrix T, readout measurements are governed by emission matrix E. \textbf{c,} (Left) Sequence of qubit readout signals for two events. (Right) Reconstructed initial cavity state probabilities. We observe an exponential suppression of the detector error based false positive probability.}
    \label{fig:pulse_meas}
\end{figure}

In order to account for all possible error mechanisms during the measurement protocol, we model the evolution of the cavity, qubit, and readout as a hidden Markov process where the cavity and qubit states are hidden variables that emit as a readout signal (see Fig. \ref{fig:pulse_meas}(b)). The Markov chain is characterized by the transition matrix (T) (Eqn. \ref{eqn:T_matrix}) that governs how the joint cavity, qubit hidden state $s \in [\ket{0g}, \ket{0e}, \ket{1g}, \ket{1e}]$ evolve, and the emission matrix (E) (Eqn. \ref{eqn:E_matrix}) which determines the probability of a readout signal R $\in$ [$\mathcal{G}$,$\mathcal{E}$] given a possible hidden state.

The transition matrix captures the possible qubit (cavity) state changes. Qubit (cavity) relaxation $\ket{e} \rightarrow \ket{g}$ ($\ket{1} \rightarrow \ket{0}$) occurs with a probability $P_{eg}^{\downarrow} = 1 - e^{-t_m/T_1^q}$ ($P_{10} = 1 - e^{-t_m/T_1^s}$). The probability of spontaneous heating $\ket{g} \rightarrow \ket{e}$ ($\ket{0} \rightarrow \ket{1}$) of the qubit (cavity) towards its steady state population is given by $P_{ge}^{\uparrow} = \bar{n}_q [1 - e^{-t_m/T_1^q}]$ ($P_{01} = \bar{n}_c [1 - e^{-t_m/T_1^s}]$). $\bar{n}_c$ is set to zero in the model in order to penalize events in which a photon appears in the cavity after the measurement sequence has begun. This makes the detector insensitive to cavity heating events. Dephasing during the parity measurement occurs with probability $P^{\phi} = 1 - e^{-t_p/T_2^q}$, leading to outcomes indistinguishable from qubit heating or decay. The transition matrix contains all qubit errors: $P_{ge} = P_{ge}^{\uparrow} + P^{\phi}$ and $P_{eg} = P_{eg}^{\downarrow} + P^{\phi}$. $P_{gg}, P_{ee}, P_{00},$ and $P_{11}$ correspond to events where no error occurs, such that probabilities pairwise sum to unity (e.g. $P_{gg} + P_{ge} = 1$). These probabilities are calculated using independently measured qubit coherences ($T_1^q = \SI{108(18)} {\micro \second}$, $T_2^q = \SI{61(4)} {\micro \second}$), cavity lifetime ($T_1^s = \SI{546(23)} {\micro \second}$), qubit spurious excited state population ($\bar{n}_q = \num{5.1(3)} \times 10^{-2}$), the length of the parity measurement ($t_p = \SI{380} {\nano \second}$), and the time between parity measurements ($t_m = \SI{10} {\micro \second}$) (see Supplemental Material for descriptions of experimental protocols used to determine these parameters \cite{Pechal_2014, Kurpiers_2018, Rosenblum_2018, Magnard_2018, Jin_2015}). The repetition rate of the experiment is constrained primarily by the readout time ($\SI{3} {\micro \second}$) and time for the readout resonator to relax back to the ground state.

\begin{equation}
    T =
    \begin{blockarray}{ccccc}
    \ket{0g} & \ket{0e} & \ket{1g} & \ket{1e} \\
    \begin{block}{[cccc]c}
    P_{00}P_{gg} & P_{00}P_{ge} & P_{01}P_{ge} & P_{01}P_{gg} & \hspace{3pt} \ket{0g} \\
    P_{00}P_{eg} & P_{00}P_{ee} & P_{01}P_{ee} & P_{01}P_{eg} & \hspace{3pt} \ket{0e}\\
    P_{10}P_{gg} & P_{10}P_{ge} & P_{11}P_{ge} & P_{11}P_{gg} & \hspace{3pt} \ket{1g}\\
    P_{10}P_{eg} & P_{10}P_{ee} & P_{11}P_{ee} & P_{11}P_{eg} & \hspace{3pt} \ket{1e}\\
    \end{block}
    \end{blockarray}
    \label{eqn:T_matrix}
\end{equation}

The elements of the emission matrix are composed of the readout fidelities of the ground and excited states of the qubit ($F_{g\mathcal{G}} = \SI{95.8(4)} {\percent}, F_{e\mathcal{E}} = \SI{95.3(5)} {\percent}$). Noise from the first stage cryogenic HEMT amplifier sets the readout fidelity.

\begin{equation}
    E = \frac{1}{2} \hspace{3pt}
    \begin{blockarray}{ccc}
    \mathcal{G} & \mathcal{E} \\
    \begin{block}{[cc]c}
    F_{g\mathcal{G}} & F_{g\mathcal{E}} & \hspace{3pt} \ket{0g}\\
    F_{e\mathcal{G}} & F_{e\mathcal{E}} & \hspace{3pt} \ket{0e}\\
    F_{g\mathcal{G}} & F_{g\mathcal{E}} & \hspace{3pt} \ket{1g}\\
    F_{e\mathcal{G}} & F_{e\mathcal{E}} & \hspace{3pt} \ket{1e}\\
    \end{block}
    \end{blockarray}
    \label{eqn:E_matrix}
\end{equation}

Given a set of $N+1$ measured readout signals ($R_0, R_1, ..., R_N$), we reconstruct the initial cavity state probabilities $P(n_0=0)$ and $P(n_0=1)$ by using the backward algorithm (Eqn. \ref{eqn:backward}) \cite{Zheng_2016, Hann_2018} and summing over all possible initial qubit states.

\begin{equation}
\begin{aligned}
    P(n_0) = \sum_{s_0 \in [\ket{n_0,g}, \ket{n_0,e}]} \sum_{s_1} ... \sum_{s_N} & E_{s_0,R_0} T_{s_0,s_1} E_{s_1,R_1} \\
    ... & T_{s_{N-1},s_N} E_{s_N,R_N}
    \label{eqn:backward}
\end{aligned}
\end{equation}

This reconstruction includes terms corresponding to all the possible processes that could occur. For example, a readout measurement of $\mathcal{G}$ followed by $\mathcal{E}$ could occur due the correct detection of a photon in the cavity (with probability $P_{11}P_{gg}F_{e\mathcal{E}}/2$). Alternatively, this measurement could be produced by a qubit heating event ($P_{00}P_{ge}F_{e\mathcal{E}}/2$) or a readout error ($P_{00}P_{gg}F_{g\mathcal{E}}/2$). Fig. \ref{fig:pulse_meas}(c) displays the measured readout signals and reconstructed initial cavity probabilities of two events. The top panels correspond to the absence of a cavity photon and the bottom panels indicate the presence of a photon.

We apply a likelihood ratio test $(\lambda =\frac{P(n_0=1)}{P(n_0=0)})$ to the reconstructed cavity state probabilities to determine if the cavity contained zero or one photons. If the likelihood ratio is greater than (less than) a threshold, $\lambda > \lambda_{\mathrm{thresh}}$ ($\lambda \leq \lambda_{\mathrm{thresh}}$), we determine the cavity to contain one (zero) photon. The probability of a detector error induced false positive is therefore less than $\frac{1}{\lambda_{\mathrm{thresh}}+1}$. As the threshold for detection increases, so too does the number of repeated parity measurements needed to confirm the presence of a photon, exacting a cost to detection efficiency that is linear in the number of measurements. More importantly for the detection of rare events, false positives are exponentially suppressed with more repeated measurements, as evident in Fig. \ref{fig:pulse_meas}(c).

\section {Detector characterization}

\begin{figure}[t!]
    \centerline{
    \includegraphics[width=\columnwidth]{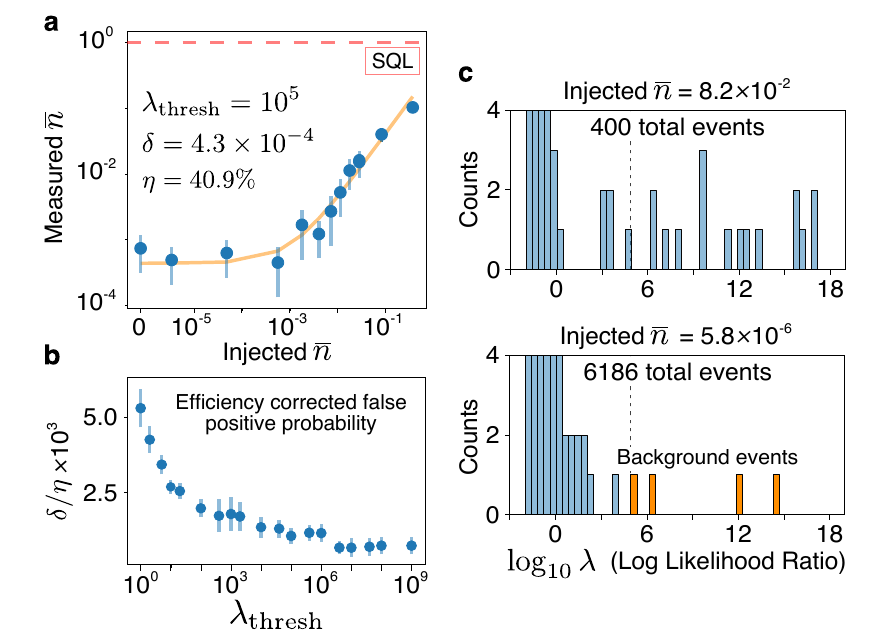}
    }
    \caption{\textbf{Detector characterization}. \textbf{a,} After a variable initial cavity displacement, 30 repeated parity measurements of cavity photon state are performed and a threshold $\lambda_{\mathrm{thresh}}$ is applied to determine the cavity population. Detector efficiency ($\eta$) and false positive probability ($\delta$) are determined from the fit in orange. The dashed red line corresponds to the standard quantum limit, which results in the noise-equivalent of one photon occupation. \textbf{b,} The efficiency corrected false positive probability ($\delta/\eta$) vs threshold ($\lambda_{\mathrm{thresh}}$) curve asymptotes at high thresholds, indicating qubit errors are now a subdominant contribution to the total detector false positive probability. \textbf{c,} Histograms of log likelihood ratios of all events for two different injected mean photon numbers. The histogram y-axis is cut off at 4 counts to view the rare events at high log likelihood ratios. The dashed grey line corresponds to $\lambda_{\mathrm{thresh}} = 10^5$ used in \textbf{a}. The unexpected photon events when very small photon numbers are injected with log likelihood ratios are from a photon background occupying the storage cavity rather than detector error based false positives.}
    \label{fig:photon_count_hist}

\end{figure}

To characterize the detector, we populate the cavity by applying a weak drive ($\bar{n} \ll 1$). We map out the relationship between the probability of injected and measured photons (Fig. \ref{fig:photon_count_hist}(a)) by varying the injected mean photon population ($\bar{n} = \alpha^2$), performing 30 repeated parity measurements, and applying $\lambda_{\mathrm{thresh}}$ to discriminate between one and zero photon events. We fit this relationship with the function $\bar{n}_{\mathrm{meas}} = \eta \bar{n}_{\mathrm{inj}} + \delta$. We obtain the efficiency of detection $\eta = \num{0.409(55)}$ and the false positive probability $\delta = \num{4.3(11)} \times 10^{-4}$ at threshold $\lambda_{\mathrm{thresh}} = 10^5$ with goodness of fit $\chi_{\mathrm{fit}}^2 = 0.0048$. 

Fig. \ref{fig:photon_count_hist}(b) shows the efficiency corrected false positive probability ($\delta/\eta$) initial decrease for low likelihood thresholds $\lambda_{\mathrm{thresh}}$, indicating a suppression of qubit and readout based false positives. Leveling off at larger thresholds indicates that the dominant source of false positives is no longer detector errors, but rather a background of real photons.

False positives that occur when qubit errors are highly suppressed (at large $\lambda_{\mathrm{thresh}}$) are due to a photon background in the storage cavity. In experiments with no photons injected into the cavity, we observe events with high likelihood ratios comparable with those seen in experiments with injected photons (Fig. \ref{fig:photon_count_hist}(c)). The detector thus correctly identifies real photons which set the background for dark matter searches. We measure the background cavity occupation to be $\bar{n}_c = \num{7.3(29)} \times 10^{-4}$, corresponding to a temperature of $\SI{39.9(22)} {\milli \kelvin}$.

Because the measured cavity photon temperature is greater than the physical $\SI{8}{\milli \kelvin}$ temperature of the device there must be coupling to extraneous baths. One contribution, arising from coupling to quasiparticles via qubit dressing of the cavity \cite{Serniak_2018}, results in a photon population of $\bar{n}_c^{q} = \num{1.8(1)} \times 10^{-4}$ (see Supplemental Material). Suppression of quasiparticle production could be achieved by enhanced infrared filtering, extensive radiation shielding, gap engineering, and quasiparticle trapping \cite{Christensen_2019, Veps_2020, Riwar_2016}. Other sources of background photons could include blackbody radiation from higher temperature stages of the dilution refrigerator, poorly thermalized or insufficiently attenuated microwave lines, or amplifier noise \cite{Yeh_2017, Wang_2019}.

\section {Hidden photon dark matter exclusion}
\begin{figure}[t!]
    \centerline{
    \includegraphics[width=\columnwidth]{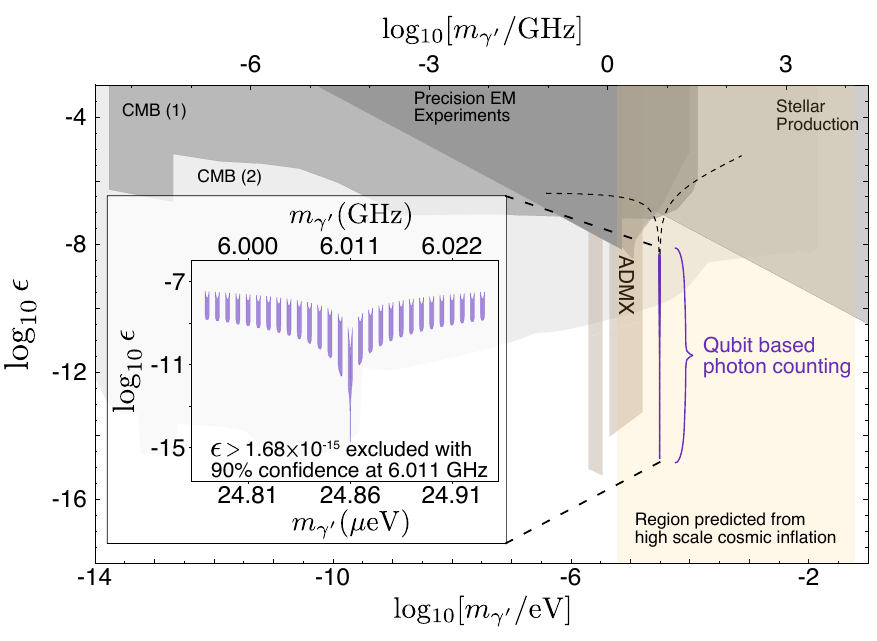}
    }
    \caption{\textbf{Hidden photon dark matter parameter space}. Shaded regions in the hidden photon parameter space \cite{Arias_2012, McDermott_2020} of coupling ($\epsilon$) and mass ($m_{\gamma}$) are excluded. In the orange band, hidden photon dark matter is naturally produced in models of high scale cosmic inflation \cite{Graham_2016}. The exclusion set with the qubit based photon counting search presented in this work, is shown in purple (dashed black line to guide the eye). On resonance with the storage cavity ($m_{\gamma '} c^2 = \hbar \omega_s$), the hidden photon kinetic mixing angle is constrained to $\epsilon \leq 1.68 \times 10^{-15}$ with 90\% confidence. The Ramsey measurement procedure is also sensitive to signals that produce cavity states with odd photon number populations greater than the measured background. Sensitivity to off resonant candidates gives rise to bands of exclusion (see inset) centered around regions where the photon number dependent qubit frequency shift is an odd multiple of $2\chi$ \cite{Gambetta_2006}. Sensitivity to large amplitude and highly detuned signals is limited by the bandwidth of the $\pi/2$ pulses used in the parity measurements.}
    \label{fig:limit_plot}
\end{figure}

By counting photons with repeated parity measurements and applying a Markov model based analysis, we demonstrate single photon detection with background shot noise reduced to $-10 \log_{10}{\sqrt{\bar{n}_c}} = \SI{15.7(9)} {\decibel}$ below the quantum limit. We use this detection technique to conduct a narrow band hidden photon search. We collect $15{,}141$ independent measurements where the injected $\bar{n}$ is well below the background population $\bar{n}_c$ and the time between measurements is much longer than either cavity or qubit timescale. Each measurement consists of integrating the signal (for the cavity lifetime, $T_1^s = \SI{546} {\micro \second}$) and counting the number of photons in the cavity with 30 repeated parity measurements ($30 \times t_m =  \SI{300} {\micro \second}$). The total search time is $15{,}141 \times (546 + 300) \si{\micro \second} = \SI{12.81} {\second}$ with a duty cycle of $\frac{546 \mu s}{846 \mu s} = 65\%$ ($\SI{8.33} {\second}$ of integration). We apply a detection threshold of $\lambda_{\mathrm{thresh}} = 10^5$, such that the qubit and readout errors are suppressed below the background photon probability ($\frac{1}{\lambda_{\mathrm{thresh}}+1} < \bar{n}_c$). We count 9 photons in $15{,}141$ measurements. Accounting for the systematic uncertainties of the experiment (statistical uncertainties are dominant, see Supplemental Material for full treatment of all systematics \cite{Conrad_2003, Rolke_2005}), a hidden photon candidate on resonance with the storage cavity ($m_{\gamma '} c^2 = \hbar \omega_s$), with mixing angle $\epsilon > 1.68 \times 10^{-15}$ is excluded at the 90\% confidence level. Fig. \ref{fig:limit_plot} shows the regions of hidden photon parameter space excluded by the qubit based search, assuming the hidden photon comprises all the dark matter density ($\rho_{\mathrm{DM}} = \SI{0.4} {\giga \electronvolt / \centi \meter^3}$). The detector is maximally sensitive to dark matter candidates with masses within a narrow window around the resonance frequency of the cavity. This window is set by the lineshape of the dark matter \cite{Foster_2018} ($Q_{\mathrm{DM}} \sim 10^6$) such that the sensitivity falls to half the maximum (-3dB point) $\SI{3} {\kilo \hertz}$ away from the cavity resonance. Additionally, sensitivity to off resonant candidates occurs in regions where the photon number dependent qubit shift is an odd multiple of the dispersive shift $2\chi$ (see Supplemental Material for calculation of hidden photon constraints \cite{Gambetta_2006}).

\section {Conclusions}
Photon number measurements allow us to gain unprecedented sensitivity to dark matter signals. The single photon counting protocol demonstrated in this work results in a $\SI{15.7} {\decibel}$ metrological gain, relative to the SQL. This improvement is currently limited by background photons $\bar{n}_c = \num{7.3} \times 10^{-4}$ whose suppression by improved filtering and shielding will further increase detector sensitivity.

In a full scale dark matter search, where the cavity is tuned to scan a wide range of dark matter masses, it is possible to estimate and subtract the background population of the cavity. The standard technique is to measure the photon population as the cavity is tuned to neighboring cavity frequencies separated by more than the dark matter linewidth. The signal hypothesis can be tested by repeating the experiment with an auxiliary cavity of the same frequency as the detection cavity, but with poor coupling to the dark matter.

The integration time required for a background limited dark matter search is determined by the signal rate ($R_s = \bar{n}_{\mathrm{DM}}/T_1^s$) and background rate ($R_b = \bar{n}_c/T_1^s$) : $R_s t > \sqrt{R_b t}$. The signal integration time scales with the background photon probability: $t > \bar{n}_c T_1^s/\bar{n}^2_{\mathrm{DM}}$. The photon detection technique developed in this work constitutes a $\bar{n}_{\mathrm{SQL}}/\bar{n}_c \sim 1300$ times speed up of dark matter searches, relative to a linear quantum limited amplifier.

This unprecedented sensitivity enables future cavity based searches for axions and hidden photons in the 3-30 GHz range. At lower frequencies, thermal backgrounds will dominate and at higher frequencies near the aluminium Josephson junction plasma frequency, qubit losses will degrade the measurement. A fixed frequency qubit can be coupled to a tunable cavity to scan over a dark matter mass range of order $\mathcal{O}(\mathrm{GHz})$, limited by the tuning range of the cavity. As long as the photon number dependent shift $2\chi$ is resolvable and the qubit and cavity are sufficiently detuned at each tuning, the QND counting protocol can be harnessed to perform a search with sub-SQL noise. A nonlinear element made of higher $T_c$ superconductor, such as tantalum \cite{Place_2020}, niobium, or titanium nitride, could be used to access frequencies beyond $\SI{30} {\giga \hertz}$ (see Supplemental Material for more information about future dark matter searches \cite{Alesini_2020, McClure_2016, Walter_2017, Leung_2018, Axline_2018, Chakram_2020_b}).

High fidelity non destructive photon counting can be utilized for accurate primary thermometry in low temperature microwave systems. This technique is applicable to quantum computing architectures which utilize long lived storage cavities \cite{Naik_2017, Gao_2018}. Assessing the residual cavity population independently of the qubit errors allows for both single shot and real time monitoring of the storage cavity, crucial when preparing states whose fidelity is sensitive to the initial conditions.

In this work, we demonstrate a state of the art photon counter for dark matter sensing. More generally, this technique of performing many QND measurements within a mode resolution time can be used more to perform sub-SQL metrology in other quantum sensing applications.

\section{Acknowledgements}
We thank N. Earnest, A. Oriani, and C. Hann for discussions. We gratefully acknowledge the support provided by the Heising-Simons Foundation. This work made use of the Pritzker Nanofabrication Facility of the Institute for Molecular Engineering at the University of Chicago, which receives support from Soft and Hybrid Nanotechnology Experimental (SHyNE) Resource (NSF ECCS-1542205), a node of the National Science Foundation’s National Nanotechnology Coordinated Infrastructure. This manuscript has been authored by Fermi Research Alliance, LLC under Contract No. DE-AC02-07CH11359 with the U.S. Department of Energy, Office of Science, Office of High Energy Physics, with support from its QuantISED program. We acknowledge support from the Samsung Advanced Institute of Technology Global Research Partnership.

\newpage


\renewcommand{\thesection}{\Alph{section}}
\renewcommand{\thefigure}{S\arabic{figure}}
\setcounter{figure}{0}
\renewcommand{\thetable}{S\arabic{table}}
\setcounter{table}{0}
\renewcommand{\theequation}{S\arabic{equation}}


\onecolumngrid
\begin{center}
    \vspace{5ex}
    \centerline{\large \textbf{Searching for Dark Matter with a Superconducting Qubit}}
    \vspace{3ex}
    \centerline{\large \textbf{Supplemental Material}}
    \vspace{3ex}
    \normalsize Akash V. Dixit, Srivatsan Chakram, Kevin He, Ankur Agrawal, Ravi K. Naik, \\ David I. Schuster, Aaron Chou
    \vspace{5ex}
\end{center}

\twocolumngrid


\section{Dark matter induced signal}
We use the potential interaction of the dark matter with electromagnetism as the basis for a search protocol. The dark matter candidate forms an effective oscillating current density that sources Maxwell's equations. Via Faraday's law, the electric field of a microwave cavity is sourced by the effective current formed by the dark matter $\nabla \times \textbf{B} -\frac{\partial \textbf{E}}{\partial t} = \textbf{j}_{\mathrm{DM}}$. For axions, the effective current density is $\textbf{j}_{\mathrm{axion}} = g_{a \gamma \gamma} \sqrt{2 \rho} \textbf{B}_0 e^{i m_a t}$, where $g_{a \gamma \gamma}$ is the predicted coupling of the axion field to electromagnetism, $\rho$ is the local dark matter density, $\textbf{B}_0$ is a DC magnetic field applied in the laboratory, and $m_a$ is the mass of the axion. For hidden photons, the effective current is $\textbf{j}_{\mathrm{HP}} = \epsilon m_{\gamma'} \sqrt{2\rho} e^{i m_{\gamma'} t} \hat{\textbf{u}}$, where $\epsilon$ is a postulated kinetic angle of mixing between standard electromagnetism and hidden sector electromagnetism, $\hat{\textbf{u}}$ is the polarization of the hidden photon field, and $m_{\gamma'}$ is the hidden photon mass. A microwave cavity tuned to the hypothetical mass of the dark matter candidate is used to accumulate the signal before it is read out.

\section{Flute cavity fabrication}
\begin{figure}[hbt!]
    \centering

    \includegraphics[width=\columnwidth]{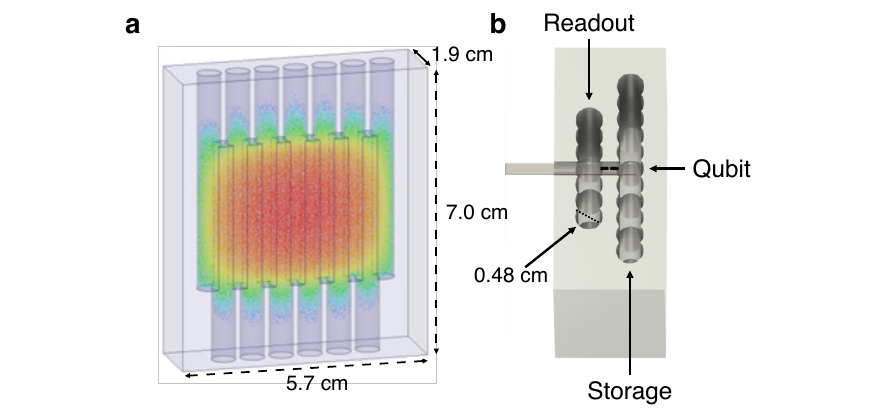}
    \caption{\textbf{Illustration of the device.} \textbf{a,} A monolithic rectangular flute cavity composed entirely of blind evanescent holes with diameter equal to the width of the cavity, drilled from both the top and bottom of the stock. The depth of the evanescent holes is chosen so the exponentially decaying field results in quality factors $> 10^9$. The electric field ($\left| \mathbf{E} \right|$) of its fundamental mode is plotted on a logarithmic scale. \textbf{b,} Rendering of half of the device including storage cavity, readout cavity, and transmon qubit.}
    \label{fig:flute}
\end{figure}

The cavities used in this work are fabricated from high purity (99.9999\%) Aluminium using a novel flute method illustrated in Fig. \ref{fig:flute}. This technique involves drilling offset holes from the top and bottom of the stock material, with a region of overlap defining the cavity volume \cite{Chakram_2020}. Making the cavity from a monolithic piece of Aluminium eliminates seam loss by design \cite{Lei_2020} and results in a high quality factor. The full device consists of two microwave cavities each coupled to the transmon qubit. One cavity has a long lived storage mode and the other is strongly coupled to the line to perform qubit readout.

\section{Transmon qubit fabrication}
The transmon qubits were fabricated on $\SI{430}{\micro \meter}$ thick C-plane (0001) Sapphire wafers with a diameter of $\SI{50.8} {\milli \meter}$. Wafers were cleaned with organic solvents (Toluene, Acetone, Methanol, Isopropanol, and DI water) in an ultrasonic bath to remove contamination, then were annealed at $\SI{1200} {\degreeCelsius}$ for 1.5 hours. Prior to film deposition, wafers underwent a second clean with organic solvents (Toluene, Acetone, Methanol, Isopropanol, and DI water) in an ultrasonic bath.  The base layer of the device, which includes the capacitor pads for the transmon, consists of $\SI{75} {\nano \meter}$ of Nb deposited via electron-beam evaporation at $\SI{1} {\angstrom / \second}$. The features were defined via optical lithography using AZ MiR 703 photoresist, and exposure with a Heidleberg MLA150 Direct Writer. The resist was developed for 1 minute in AZ MIF 300 1:1. The features were etched in a Plasma-Therm inductively coupled plasma (ICP) etcher using fluorine based ICP etch chemistry with a plasma consisting of 15 sccm $\mathrm{SF_6}$, 40 sccm $\mathrm{CHF_3}$, and 10 sccm $\mathrm{Ar}$. The junction mask was defined via electron-beam lithography of a bi-layer resist (MMA-PMMA) in the Manhattan pattern using a Raith EBPG5000 Plus E-Beam Writer, with overlap pads for direct galvanic contact to the optically defined capacitors. The resist stack was developed for 1.5 minutes in a solution of 3 parts IPA and 1 part DI water. Before deposition, the overlap regions on the pre-deposited capacitors were milled \textit{in-situ} with an Argon ion mill to remove the native oxide. The junctions were then deposited with a three step electron-beam evaporation and oxidation process. First, an initial $\SI{35} {\nano \meter}$ layer of aluminium was deposited at $\SI{1} {\nano \meter / \second}$ at an angle of $\SI{29} {\degree}$ relative to the normal of the substrate, parallel azimuthally to one of the fingers in the Manhattan pattern for each of the junctions. Next, the junctions were exposed to $\SI{20} {\milli \bar}$ of a high-purity mixture of $\text{Ar}$ and $\text{O}_2$ (ratio of 80:20) for 12 minutes for the first layer to grow a native oxide. Finally, a second $\SI{120} {\nano \meter}$ layer of aluminium was deposited at $\SI{1} {\nano \meter / \second}$ at the same angle relative to the normal of the substrate, but orthogonal azimuthally to the first layer of aluminium. After evaporation, the remaining resist was removed via liftoff in N-Methyl-2-pyrrolidone (NMP) at $\SI{80} {\degreeCelsius}$ for 3 hours, leaving only the junctions directly connected to the base layer. After both the evaporation and liftoff, the device was exposed to an ion-producing fan for 15 minutes, in order to avoid electrostatic discharge of the junctions. The room temperature DC resistance of the Josephson junction on each qubit was measured to select the qubit which corresponds to the target Josephson energy \cite{Ambegaokar_1963} ($E_J$).

\section{Experimental setup}
\begin{figure}[hbt!]
    \centering
    \includegraphics[width=\columnwidth]{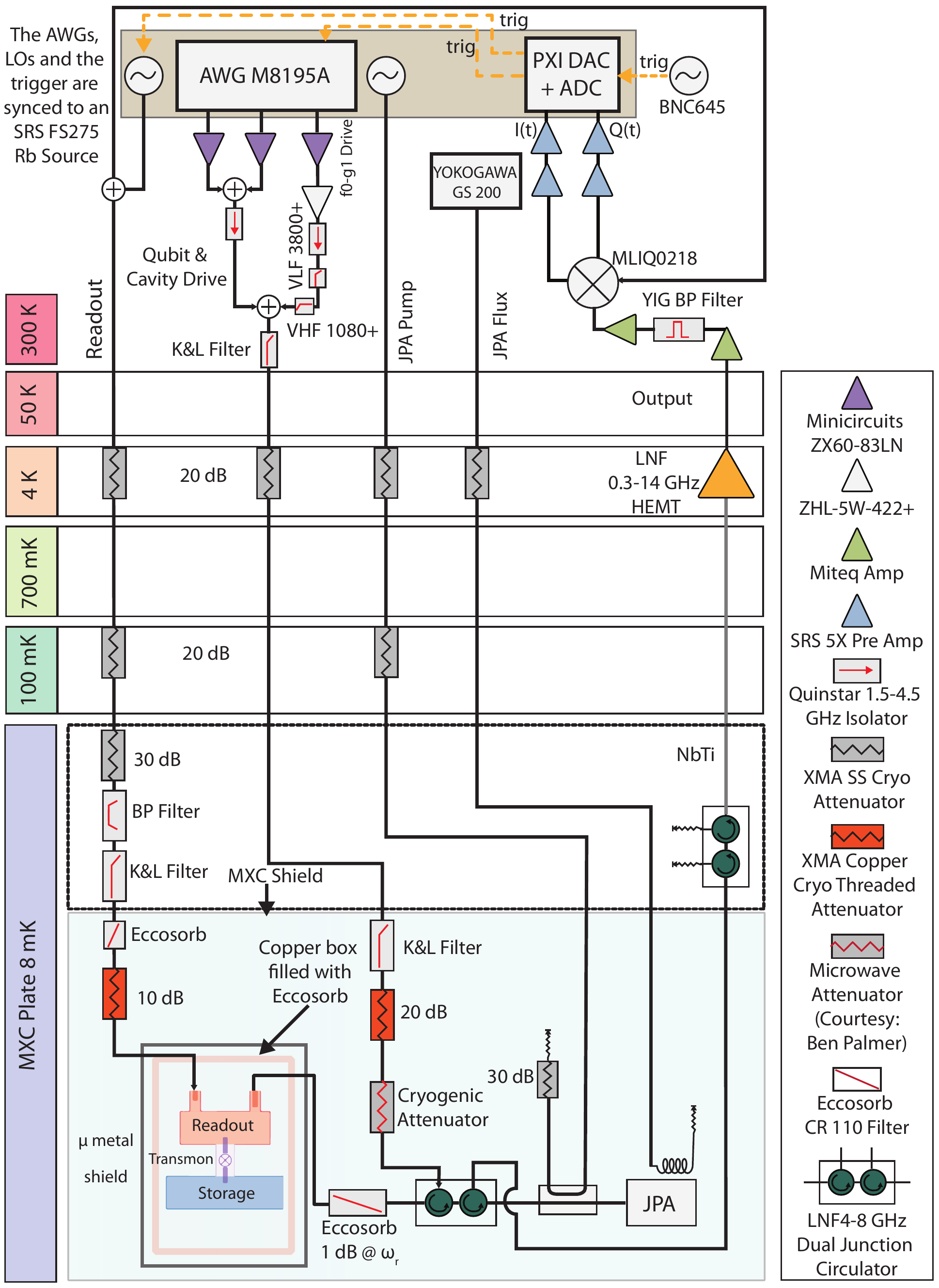}
    \caption{\textbf{Wiring diagram inside the dilution refrigerator and the room temperature measurement setup.} Qubit readout is performed by injecting a drive into the weakly coupled port. After interacting with the readout cavity, the signal is routed to the amplification chain using non reciprocal circulator and isolator elements. We note, the Josephson parametric amplifier is not in operation for the measurements presented in this work. The signal is then mixed down to DC, further amplified, and finally digitized. Qubit and storage cavity operations are performed via the strongly coupled port. This line is heavily filtered and attenuated \cite{Yeh_2017} to minimize stray radiation from entering the device.}
    \label{fig:wiring_diagram}
\end{figure}

The cavities and qubit are mounted to the base plate of a dilution fridge (Bluefors LD400) operating at $\SI{8} {\milli \kelvin}$. The device is potted in a block of infrared (IR) absorbant material (eccosorb CR-110) to absorb stray radiation and housed in two layers of $\mu$-metal to shield from magnetic fields. Signals sent to the device are attenuated and thermalized at each temperature stage of the cryostat as shown in Fig. \ref{fig:wiring_diagram}. The field probing the readout resonator is injected via the weakly coupled port (shorter dipole stub antenna). Control pulses for qubit, storage cavity, and sideband operation are inserted through the strongly coupled readout port (longer dipole stub antenna). This line includes a cryogenic microwave attenuator thermalized to the base stage (Courtesy of B. Palmer) and a weak eccosorb (IR filter). Both control lines also contain an inline copper coated XMA attenuator that is threaded to the base state. The signal from the readout resonator reflects off a Josephson parametric amplifier (not used in this work) before being amplified by a cryogenic HEMT amplifier at the $\SI{4} {\kelvin}$ stage. The output is mixed down to DC before being digitized.

\section{Calibration of parity measurement}
The cavity number parity measurement requires the calibration of the two qubit $\pi/2$ pulses as well as the delay between them. To set the $\pi/2$ pulse length, we perform qubit Rabi oscillations between the $\ket{g}$ and $\ket{e}$ levels by driving at the qubit transition frequency. The population transfer is sinusoidal and can be fit to determine when the qubit population has inverted ($\pi$ pulse). By turning on the drive for only half the time required to perform a $\pi$ pulse gives us the needed $\pi/2$ pulse to put the qubit in a clock state $(\frac{1}{\sqrt{2}} (\ket{g} + \ket{e}))$. The parity measurement is comprised of an initial $\pi/2$ pulse followed by a time delay of $\pi/ |2\chi|$ and a final $-\pi/2$ pulse (constructed by advancing the phase of the $\pi/2$ pulse by $\pi$).

To calibrate the time delay used in the parity measurement, we perform two Ramsey interferometry experiments on the qubit either in the absence or presence of a single photon in the cavity. We chose the Ramsey drive frequency to be on resonance with the qubit transition frequency. In the absence of the photon, the qubit superposition remains unchanged in the frame of the qubit. We use the $\ket{f0}-\ket{g1}$ sideband \cite{Pechal_2014, Kurpiers_2018, Rosenblum_2018, Magnard_2018} to populate the cavity with a single photon. In the presence of the cavity photon, the qubit transition frequency is shifted by $2 \chi$ relative to the Ramsey  frequency, consequently, the resulting fringe oscillates at a rate of $2 \chi$. The parity measurement delay time ($t_p$) is chosen such that the qubit superposition state has obtained a phase shift of $\pi$. This is also a calibration of the dispersive shift ($|2 \chi| = \pi/t_{p}$).

\section{Drive pulse calibration}
\begin{figure}[hbt!]
    \centerline{
    \includegraphics[width=\columnwidth]{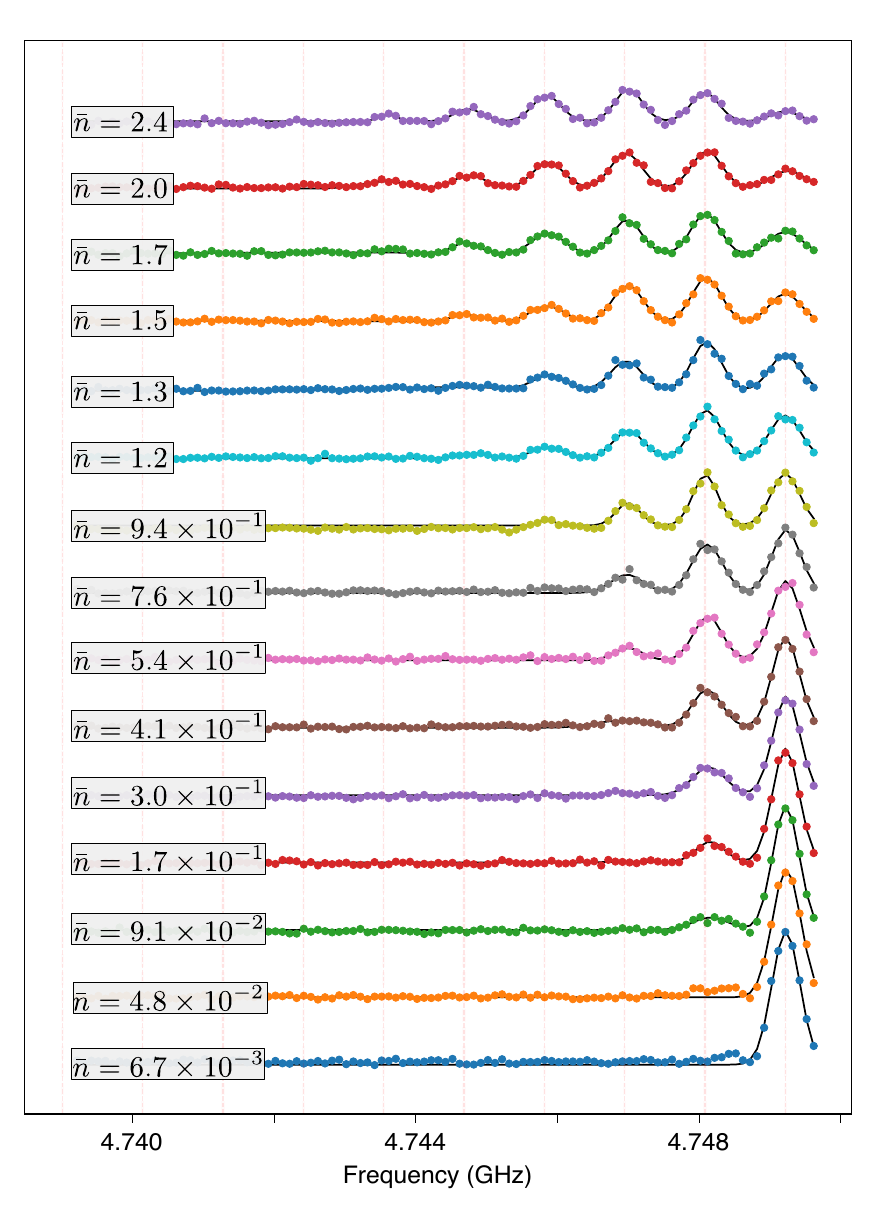}
    }
    \caption{\textbf{Qubit spectroscopy reveals cavity displacement} The cavity is displaced using a variable weak coherent drive for a finite period of time. The resulting population of the cavity is determined by performing qubit spectroscopy (points). The cavity photon number dependent shift of the qubit transition frequency reveals the cavity population. By fitting to the spectrum (black) we extract the weights of the cavity number states in the prepared coherent state.}
    \label{fig:alpha_cal}
\end{figure}

By applying a weak coherent tone at the storage cavity frequency, we induce a variable displacement $\alpha$ of the cavity state. We calibrate the number of photons injected into the storage cavity by varying the drive amplitude and performing qubit spectroscopy. By fitting the qubit spectrum show in Fig. \ref{fig:alpha_cal} to a Poisson distribution, we extract the cavity occupation, $\bar{n} = \mathrm{|\alpha|^{2}}$.

\begin{figure}[hbt!]
    \centerline{
    \includegraphics[width=\columnwidth]{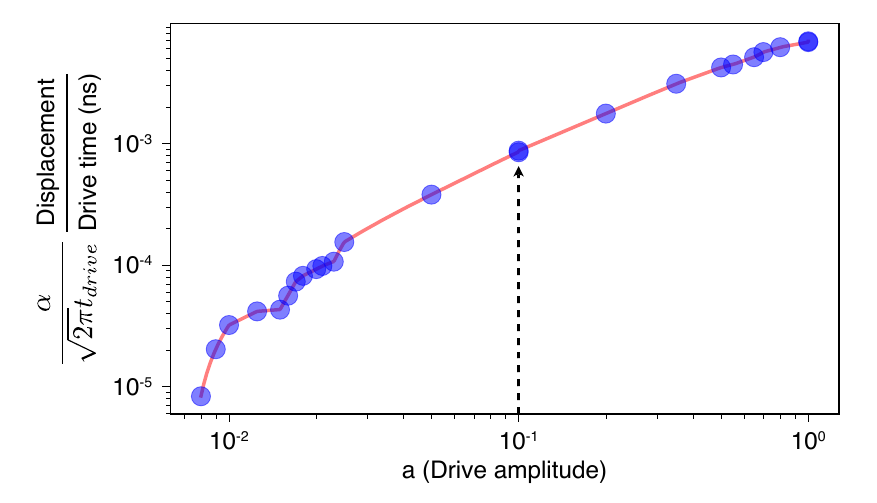}
    }
    \caption{\textbf{Transfer function describing mapping between drive amplitude in software and cavity displacement.} This transfer function is calibrated such that the cavity is displaced by $\alpha$ when we use a coherent drive of length $t_{drive}$ and amplitude of a (in software) at the cavity frequency. Blue points are obtained by fitting to qubit spectroscopy after applying cavity displacements with variable drive time. For example, a 10ns pulse with a = 0.1 (indicated by the arrow) produces a cavity displacement of $\alpha = 2.1 \times 10^{-1}$. The red curve is a linear interpolation between the data points and can be used to generate displacements that are not directly calibrated. The data points are chosen to capture the nonlinear behavior of the waveform generator at values where an additional bit is necessary to represent the drive amplitude.}
    \label{fig:drive_transfer}
    
\end{figure}

Nonlinearities of the signal generator result in a non trivial relationship between drive amplitude (at the software level) and the cavity occupation number. We map the transfer function that describes this relation and use it to apply calibrated cavity displacements (Fig. \ref{fig:drive_transfer}).

\section{Elements of hidden Markov model}
The hidden Markov model relies on independent measurements of the probabilities contained in the transition and emission matrices. The elements of these matricies depend on the parameters of the experiment and the device, including the lifetimes of the qubit and cavity, qubit spurious population, and readout fidelities.

\subsection{Transmission matrix elements}
The lifetime of the qubit is determined by applying a $\pi$ pulse and waiting for a variable time before measuring the population. We map out the qubit population as a function of the delay time, fit it with an exponential characterizing the Poissonian nature of the decay process, and obtain $T_1^q = \SI{108(18)} {\micro \second}$.

The dephasing time of the qubit is measured by a Ramsey interferometry experiment with a $\pi/2$ pulse, variable delay, and a final $\pi/2$ with its phase advanced by $\omega_r t$ where $\omega_r$ is the Ramsey frequency. During the variable delay period, a series of $\pi$ pulses are applied to perform spin echos and reduce sensitivity to low frequency noise. We observe a dephasing time of $T_2^q = \SI{61(4)} {\micro \second}$.

\begin{figure}[hbt!]
    \centerline{
    \includegraphics[width=\columnwidth]{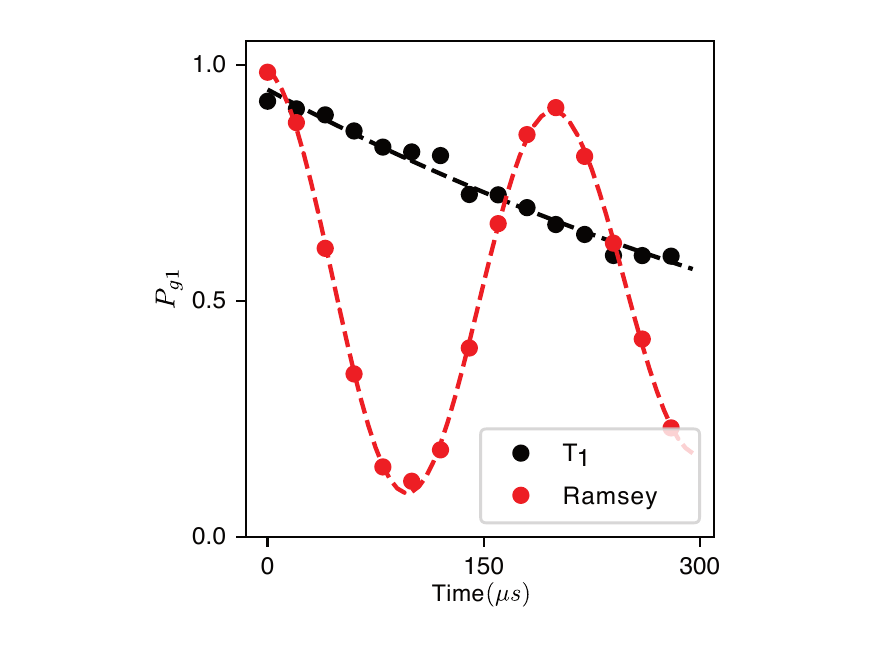}\
    }
    \caption{\textbf{Storage cavity lifetime and dephasing time from $\mathrm{T_1}$ and Ramsey measurements.} The long lived storage cavity mode is ideal for holding a signal photon induced by the dark matter while a series of repeated photon counting measurements is performed.}
    \label{fig:cavity_coherence}
\end{figure}

The storage cavity lifetime is calibrated by performing a cavity $T_1$ experiment. This is accomplished by applying a $\pi_{ge}$ pulse and a $\pi_{ef}$ to the transmon. This is followed by driving the $\ket{f0}-\ket{g1}$ transition, mediated by the Josephson nonliniearity for a time corresponding to a $\pi$ pulse \cite{Pechal_2014, Kurpiers_2018, Rosenblum_2018, Magnard_2018}. This populates the cavity with $\ket{n}=\ket{1}$ photons. After a variable time delay, the cavity population is swapped back into the qubit using the same $\pi_{\ket{f0}-\ket{g1}}$ pulse. Measuring the qubit population, we infer the cavity population as a function of the time delay. This is fit with a decaying exponential to obtain $T_1^s = \SI{546(23)} {\micro \second}$ (Fig. \ref{fig:cavity_coherence}). To measure the cavity dephasing time, the cavity is initialized in a superposition state $\frac{1}{\sqrt{2}} (\ket{0} + \ket{1})$ by first applying a $\pi_{ge}/2$ pulse, a $\pi_{ef}$, followed by a $\pi_{\ket{f0}-\ket{g1}}$ pulse. A Ramsey measurement is performed to obtain a cavity dephasing time of $T_2^s = \SI{774(286)} {\micro \second}$.

The qubit spurious population is determined by measuring the relative populations of its ground and excited states \cite{Jin_2015}. This is done by utilizing the $f$-level of the transmon. Two Rabi experiments are conducted swapping population between the $\ket{e}$ and $\ket{f}$ levels. First, we apply a $\pi_{ge}$ pulse to invert the qubit population followed by the $\ket{e}-\ket{f}$ Rabi experiment. Second, no $\pi_{ge}$ pulse is applied before the $ef$ Rabi oscillation. The ratio of the amplitudes of the oscillations gives us the ratio of the populations of the excited and ground state. Assuming that $P(g) + P(e) = 1$ and measuring $\frac{P(e)}{P(g)}$, we obtain $P(g) = 0.949$ and $P(e) = 0.051$, corresponding to an effective qubit temperature of $\SI{71} {\milli \kelvin}$.

\subsection{Emission matrix elements}
\begin{figure*}[hbt!]
    \centerline{
    \includegraphics[width=2 \columnwidth]{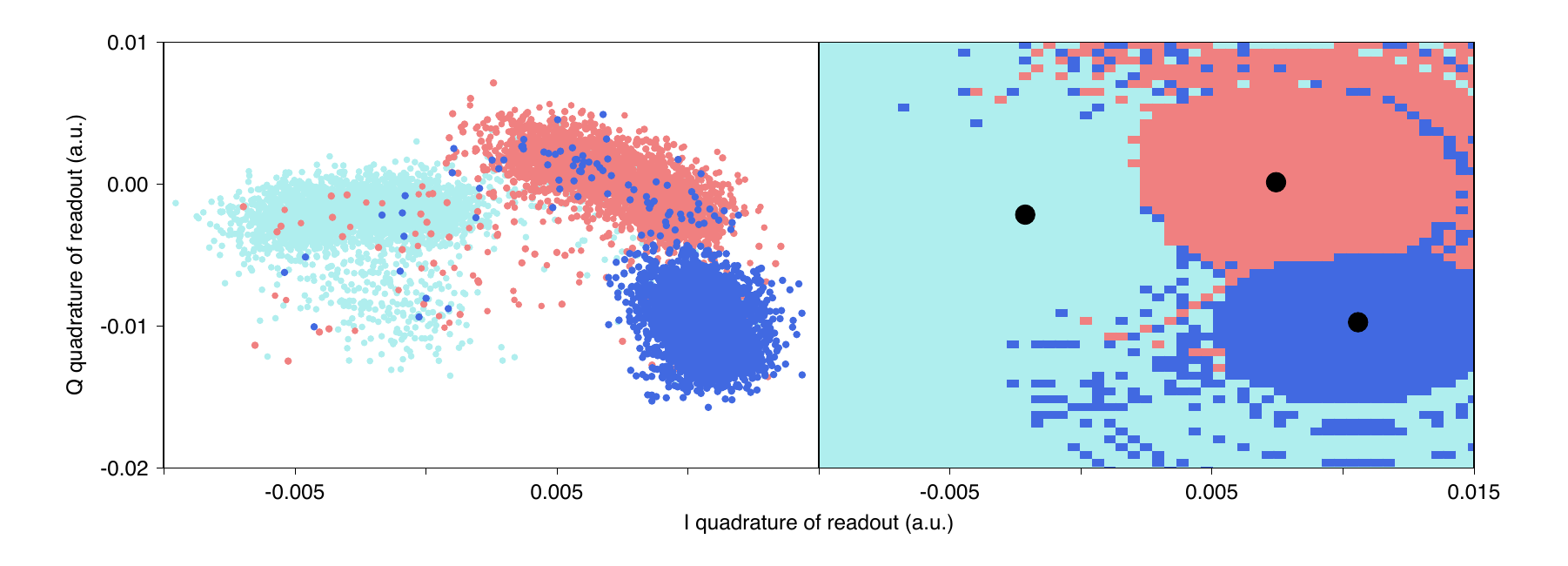}
    }
    \caption{\textbf{Mapping of readout measurements to transmon states.} (Left) Transmon is prepared in one of its possible states ($\ket{g}, \ket{e}, \ket{f}$ in blue, red, cyan) 3000 times each and the corresponding readout signals are recorded. (Right) From the readout data we can generate a map that takes a new measurement (point in IQ space) and returns a readout signal, $\mathcal{G}, \mathcal{E},$ or $\mathcal{F}$. All measurements that fall outside of the subspace of $\ket{g}$ and $\ket{e}$ are assigned to $\ket{f}$ since the parity measurement only makes use of the first two levels of the transmon.}
    \label{fig:readout_hist_map}
    
\end{figure*}

In order to characterize the emission matrix it is necessary to measure the readout infidelity of a particular transmon state. All the possible transmon states ($\ket{g}, \ket{e}, \ket{f}$) are prepared (3000 independent experiments per state) and the resulting I,Q signals are digitized. The resulting distributions in I,Q space are used as a map to determine the probability that any readout signal is the result of transmon being in either $\ket{g}, \ket{e},$ or $\ket{f}$. Based on the calculated probability, the state is assigned to either $\mathcal{G}, \mathcal{E},$ or $\mathcal{F}$ (Fig. \ref{fig:readout_hist_map}).

Readout errors are due to voltage excursions from amplifier noise or spurious qubit transitions. The emission matrix should only contain readout errors that occur due to voltage fluctuations. Errors due to qubit transitions during the readout window are accounted for in the transition matrix. To disentangle the two contributions the analysis is run with various contingencies, all resulting in the same detector false positive probability and efficiency, indicating that the Markov model is robust to small perturbations of the emission matrix when so heavily biased against false positives. In the most conservative case, the qubit errors are accounted for during the entire $\SI{10} {\micro \second}$ window or each parity measurement. The readout infidelity is determined by finding how many errors are made (regardless of the source) during the $\SI{3} {\micro \second}$ readout window (part of the $\SI{10} {\micro \second}$ experiment time window). In this case qubit errors are counted twice during the readout window. In the second case, we consider qubit errors only for times when readout is not occurring (7 of the $\SI{10} {\micro \second}$) and include all error channels in the readout infidelity. This avoids double counting of the qubit error during readout. The third case most closely aligns with the plausible physical model of errors during the readout window occurring due to a combination of qubit errors and amplifier noise. The readout infidelity is computed by subtracting the qubit error probabilities during the $\SI{3} {\micro \second}$ readout window due to qubit decay ($1-e^{-\SI{3} {\micro \second} / T_1^q}$) or heating ($\bar{n}_q[1-e^{-\SI{3} {\micro \second} / T_1^q}]$) from the total measured error during readout, leaving only readout errors due to voltage noise from amplifiers. This is the readout infidelity used to determine the elements of the emission matrix in the analysis.

\begin{table}[hbt!]
\begin{center}
\begin{tabular}{ l l }
\hline
 Device Parameter & Value \\
 \hline
 Qubit frequency & $\omega_q = 2 \pi \times \SI{4.749} {\giga \hertz}$ \\
 Qubit anharmonicity & $\alpha_q = \SI{-139.5} {\mega \hertz}$ \\
 Qubit decay time & $T_1^q = \SI{108(18)} {\micro \second}$ \\ 
 Qubit dephasing time & $T_2^q = \SI{61(4)} {\micro \second}$ \\ 
 Qubit residual occupation & $\bar{n}_q = \num{5.1(3)} \times 10^{-2}$ \\
 Storage frequency & $\omega_s = 2 \pi \times \SI{6.011} {\giga \hertz}$ \\
 Storage decay time & $T_1^s = \SI{546(23)} {\micro \second}$ \\
 Storage dephasing time & $T_2^s = \SI{774(286)} {\micro \second}$ \\
 Storage-Qubit Stark shift & $2 \chi =  - 2 \pi \times \SI{1.13} {\mega \hertz}$ \\
 Storage residual occupation &  $\bar{n}_q = \num{7.3(29)} \times 10^{-4}$ \\
 Readout frequency & $\omega_r = 2 \pi \times \SI{8.052} {\giga \hertz}$ \\
 Readout $\ket{e}$ shift & $2\chi_r^{e} =  - 2 \pi \times \SI{0.38} {\mega \hertz}$ \\
 Readout $\ket{f}$ shift & $2\chi_r^{f} =  - 2 \pi \times \SI{0.73} {\mega \hertz}$ \\
 Readout fidelity ($\ket{g}$) & $F_{g\mathcal{G}} = \SI{95.8(4)} {\percent}$ \\
 Readout fidelity ($\ket{e}$) & $F_{e\mathcal{E}} = \SI{95.3(5)} {\percent}$ \\
\hline

\end{tabular}
\caption{\textbf{Device parameters}. Measured qubit, storage, and readout cavity parameters. These independently measured values are necessary to determine for the transition and emission matrices. This enables the hidden Markov model to capture the behavior of the system during the measurement sequence.}
\label{table:device_params}
\end{center}
\end{table}

\section{Detector characterization}
To characterize the detector, the cavity population is varied by applying a weak drive and the cavity photon number is counted using the technique described in the main text. In order to extract the efficiency ($\eta$) and false positive probability ($\delta$) of the detector, the relationship between injected photon population ($\bar{n}_\mathrm{inj}$) and measured photon population ($\bar{n}_\mathrm{meas}$) is fit to $\bar{n}_\mathrm{inj} = \eta \times \bar{n}_\mathrm{meas} + \delta$.

\subsection{Detector efficiency}
\begin{figure}[hbt!]
    \centerline{
    \includegraphics[width=\columnwidth]{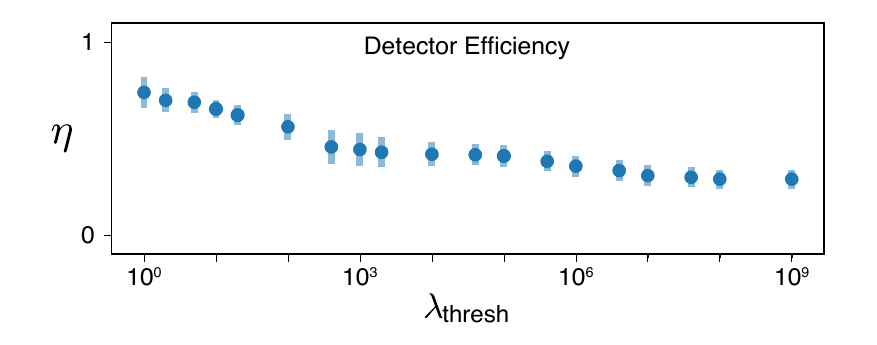}
    }
    \caption{\textbf{Detector efficiency as a function of detection threshold.} As the threshold for detection become stricter, the detector efficiency decreases. The efficiency scales linearly with the threshold, while the the false positive probability due to the detector errors decreases exponentially.}
    \label{fig:efficiency}
    
\end{figure}

The detector efficiency and false positive probability is determined at varying thresholds for detection $\lambda_{\mathrm{thresh}}$. As the detection threshold is increased, more parity measurements are required to determine the presence of a photon. This suppresses false positives due to qubit errors but also leads to a decrease in the detector efficiency as events with low likelihood ratio are now rejected (Fig. \ref{fig:efficiency}). For large thresholds where $\frac{1}{\lambda_{\mathrm{thresh}}+1} < \delta$, the qubit based errors are no longer the dominant source of detector errors. These errors occur due to the presence of a background of real photons whose population is given by the efficiency corrected false positive probability $\delta/\eta$ (shown in the main text).

\subsection{Analysis with cavity backgrounds}
In the initial application of the hidden Markov analysis, the cavity background population is set to zero to exclude the effects of cavity heating events during the measurement. We measure a background population of $\num{7.3(29)} \times 10^{-4}$ and then include $\bar{n}_c$ in the model to reanalyze the data. Now in addition to the possibility of a photon from injection or background initially occupying the cavity, we allow for both cavity heating during the measurement and qubit errors. The reconstructed false positive probability of the detector $P(n=0)$ is set by the probability of a cavity heating event preceding the first parity measurement $\bar{n}_c \times (1-e^{-t_m/T_1^s}) = 1.3 \times 10^{-5}$. Therefore, the maximum likelihood ratio of the reconstructed probabilities cannot exceed the cavity heating limit. By setting the detection threshold to $\lambda_{\mathrm{thresh}} = 2.0 \times 10^4$ and including cavity heating processes in the model, we measure the background population of the cavity to be $\num{5.6(18)} \times 10^{-4}$, which is consistent with the initial measurement of $\bar{n}_c = \num{7.3(29)} \times 10^{-4}$. This reinforces our belief that repeated parity measurements successfully mitigate qubit based errors and that cavity background photons are the limiting process for photon detection.

\section{Sources of cavity backgrounds}
The measured photon background contains a contribution from spurious photons injected due to the cavity interaction with the qubit. To characterize the significance of the effect of converted qubit excitations, we recognize that cavity ($\ket{\tilde{1}}$) and qubit ($\ket{\tilde{e}}$) excitations are dressed due to their interaction. In the dispersive limit, written in terms of the bare basis eigenstates ($\ket{e}, \ket{1}$), the dressed states are:

\begin{equation}
\begin{aligned}
    \ket{\tilde{e}} &= \sin{\theta} \ket{g,1} + \cos{\theta} \ket{e,0} \\
    \ket{\tilde{1}} &= \cos{\theta} \ket{g,1} - \sin{\theta} \ket{e,0}
\end{aligned}
\end{equation}

where $\theta$ is the mixing angle between the two systems. Qubit heating events from quasiparticle tunneling, in effect, prepare the system in the state $\ket{e,0} = \ket{\tilde{e}} - \frac{\sin{\theta}}{\cos{\theta}} \ket{\tilde{1}}$. The probability that the heating event manifests as a cavity excitation is $(\frac{\sin{\theta}}{\cos{\theta}})^2 = 3.5 \times 10^{-3}$. Therefore, the contribution of qubit heating events converted to cavity photons is determined by the probability that there is a qubit heating event and the probability that it is projected into a cavity excitation, $\bar{n}_c^q = \bar{n}_q \times (\frac{\sin{\theta}}{\cos{\theta}})^2 = \num{1.8(1)} \times 10^{-4}$.

\section{Converting cavity occupation limit to hidden photon exclusion}

\subsection{Kinetic mixing angle exclusion}

For a dark matter candidate on resonance with the cavity frequency ($m_{\mathrm{DM}} c^2 = \hbar \omega_c$), the rate of photons deposited in the cavity by the coherent build up of electric field in one cavity coherence time is given by \cite{Chaudhuri_2015}:

\begin{equation}
\frac{d N_{\mathrm{HP}}}{dt}  = \frac{U/\omega_s}{T_1^s} = \frac{1}{2} \frac{E^2 V }{\omega_s} \frac{\omega_s}{Q_s} = \frac{1}{2} \frac{J^2_{\mathrm{DM}} Q^2_{\mathrm{DM}}}{m^2} \frac{Q_s}{Q_{\mathrm{DM}}} G V \frac{1}{Q_s}
\label{eqn:N_DM}
\end{equation}
The cavity coherence time is given by $T_1^s = \frac{Q_s}{\omega_s}$. The volume of the cavity is $0.953 \times 3.48 \times \SI{3.56} {\centi \meter^3} =  \SI{11.8} {\centi \meter^3}$. $\mathcal{G}$ encompasses the total geometric factor of the particular cavity used in the experiment. This includes a factor of $1/3$ due to the dark matter field polarization being randomly oriented every coherence time. For the lowest order mode of the rectangular cavity coupled to the qubit with $\textbf{E} = \sin(\frac{\pi x}{l}) \sin(\frac{\pi y}{w})\textbf{z}$ the geometric form factor is given by:

\begin{equation}
    G = \frac{1}{3} \frac{\left| \int dV E_z \right|^2}{V \int dV \left| E_z \right|^2} = \frac{1}{3} \frac{2^6}{\pi^4}
\end{equation}

Since the cavity decay and dephasing times ($T_1^s$ and $T_2^s$) are longer than the dark matter coherence ($Q_{\mathrm{DM}} = 10^6$), the cavity is displaced $\frac{Q_s}{Q_{\mathrm{DM}}}$ times with a random phase each dark matter coherence time. The cavity field displacement follows a random walk, leading to an signal amplitude enhancement by a factor of $\sqrt{\frac{Q_s}{Q_{\mathrm{DM}}}}$.

The hidden photon generated current is set by the density of dark matter in the galaxy $\rho_{\mathrm{DM}} = \SI{0.4} {\giga \electronvolt / \centi \meter^3} = 2\pi \times \SI{9.67e19} {\giga \hertz / \centi \meter^3}$:

\begin{equation}
J^2_{\mathrm{DM}} = 2\epsilon^2 m^4 A'^2 = 2\epsilon^2 m^2 \rho_{\mathrm{DM}}
\label{eqn:J_DM}
\end{equation}

Substituting Eqn. \ref{eqn:J_DM} into Eqn. \ref{eqn:N_DM} yields the signal rate of photons deposited in the cavity by a hidden photon dark matter candidate:

\begin{equation}
\frac{d N_{\mathrm{HP}}}{dt} = \epsilon^2 \rho_{\mathrm{DM}} Q_{\mathrm{DM}} G V
\end{equation}

The total number of photons we expect to be deposited is determined by the photon rate and the integration time ($T_1^s \times N_{\mathrm{meas}} = \SI{8.33} {\second}$):

\begin{equation}
N_{\mathrm{HP}} = \frac{d N_{\mathrm{HP}}}{dt} \times T_1^s \times N_{\mathrm{meas}} = \frac{\epsilon^2 \rho_{\mathrm{DM}} Q_{\mathrm{DM}} Q_s G V N_{\mathrm{meas}}}{\omega_s}
\label{eqn:N_HP}
\end{equation}

\subsection{Calculating 90\% confidence limit}
\begin{table}[hbt!]
\begin{center}
\begin{tabular}{ l l l }
\hline
 Expt. Parameter & $\Theta$ & $\sigma_{\Theta}$ \\
 \hline
 Quantum efficiency & $\eta = 0.409$ & $\sigma_\eta=0.055$ \\
 Storage cavity frequency & $\omega_s=\SI{6.011}{\giga \hertz}$ & $\sigma_{\omega_s}=\SI{205}{\hertz}$ \\
 Storage quality factor & $Q_s = 2.06 \times 10^7$ & $\sigma_{Q_s} = 8.69 \times 10^5$ \\
 Storage cavity volume & $V=\SI{11.8} {\centi \meter^3}$ & $\sigma_V=\SI{0.2} {\centi \meter^3}$ \\
 Storage form factor & $G=0.22$ & $\sigma_G=0.003$ \\
\hline
\end{tabular}
\caption{\textbf{Experimental parameters}. Systematic uncertainties of physical parameters in the experiment must be incorporated in determining the excluded hidden photon mixing angle $\epsilon$. The uncertainty in the quantum efficiency is determined in the main text from fitting the relation between the measured and injected photon population at a detection threshold of $\lambda_{\mathrm{thresh}} = 10^5$. The storage cavity frequency uncertainty is obtained by Ramsey interferometry. The quality factor of the cavity is given by $Q_s = \omega_s T_1^s$ so the uncertainty is calculated as $\sigma_{Q_s}^2 = (\omega_s \sigma_{T_1^s})^2 + (T_1^s \sigma_{\omega_s})^2$. The volume uncertainty is estimated by assuming machining tolerances of 0.005 inches in each dimension. The form a factor uncertainty is estimated from assuming 1\% error in the simulated structure. Of the experimental quantities, the efficiency has largest fractional uncertainty (13\%), though the statistical fluctuations of the observed counts still dominate (33\%).}
\label{table:expt_params}
\end{center}
\end{table}

By counting single photons when the applied drive population less than the background population ($\bar{n}_c$) we perform a hidden photon search. We count $N = 9$ background photons in $N_{\mathrm{meas}} = 15{,}141$ measurements. We determine the hidden photon mixing angle $\epsilon$ that can be excluded at the 90\% confidence level by computing the probability that the signal could result in less than or equal to  9 photons measured ($N \leq 9$) with less than 10\% probability. In each measurement a photon is counted or not so the signal is described by a binomial distribution with probability set by the expected number of deposited photons as calculated in Eqn. \ref{eqn:N_HP}. The systematic uncertainties of the various experimentally determined quantities in Eqn. \ref{eqn:N_HP} are treated as nuisance parameters \cite{Rolke_2005} with an assumed Gaussian distribution of mean $\Theta$ and standard deviation $\sigma_{\Theta}$ as shown in Table \ref{table:expt_params}. We marginalize over the nuisance parameters \cite{Conrad_2003} and compute the cumulative probability shown in Eqn. \ref{eqn:cumul}.

\begin{equation}
\begin{aligned}
P(\leq N) = \int_{0}^{\infty} & \prod_{i} d\Theta_i' \frac{e^{-(\Theta_i-\Theta_i')^2/2\sigma_{\Theta_i}^2}}{\sqrt{2 \pi}\sigma_{\Theta_i}} \sum_{k=0}^{N}  \frac{N_{\mathrm{meas}}!}{k! (N_{\mathrm{meas}}-k)!} \\
& \times \left( \frac{\eta' \epsilon^2 \rho_{\mathrm{DM}} Q_{\mathrm{DM}} Q_s' G' V'}{\omega_s'} \right) ^{k} \\
& \times \left(1- \frac{\eta' \epsilon^2 \rho_{\mathrm{DM}} Q_{\mathrm{DM}} Q_s' G' V'}{\omega_s'} \right)^{N_{\mathrm{meas}} - k}
\label{eqn:cumul}
\end{aligned}
\end{equation}

\begin{figure}[hbt !]
    \centerline{
    \includegraphics[width=\columnwidth]{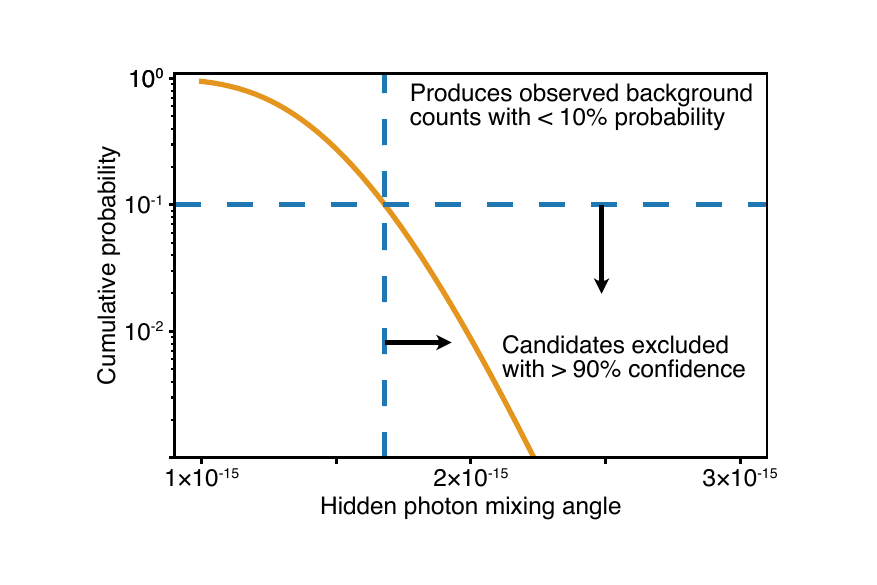}
    }
    \caption{\textbf{Cumulative probability of hidden photon candidate producing observed signal.} Regions where the cumulative probability falls below 0.1 are ruled out as potential hidden photon candidates with 90\% confidence. The minimum mixing angle that can be excluded with 90\% confidence is $1.68 \times 10^{-15}$.}
    \label{fig:eps_exclusion}
    
\end{figure}

For a given hidden photon candidate, a cumulative probability of $< 0.1$ implies that candidate has less than 10\% chance of producing the observed signal, thereby excluding such a candidate with 90\% confidence. This leads us to exclude, with 90\% confidence, hidden photon candidates with $\epsilon^{90\%} > 1.68 \times 10^{-15}$ as seen in Fig. \ref{fig:eps_exclusion}

\section{Hidden photon parameter space exclusion}
Single photon counting with repeated parity measurements is sensitive to a wide range of candidates in the parameter space of hidden photon mass ($m_{\gamma '}$) and kinetic mixing angle ($\epsilon$). To determine the sensitivity of the detector to a particular candidate, there are two considerations: the photon number dependent shift of the qubit transition as a function of the hidden photon mass, and the probability that a candidate would result in the measurement of a photon with probability larger than excluded. The photon population excluded as the 90\% confidence level is computed using the excluded mixing angle $\epsilon^{90\%}$ and Eqn. \ref{eqn:N_HP} as $\bar{n}_{\mathrm{HP}}^{90\%} = \frac{N_{\mathrm{HP}}^{90\%}}{N_{\mathrm{meas}}} = 2.42 \times 10^{-3}$.

\begin{figure}[hbt!]
    \centerline{
    \includegraphics[width=\columnwidth]{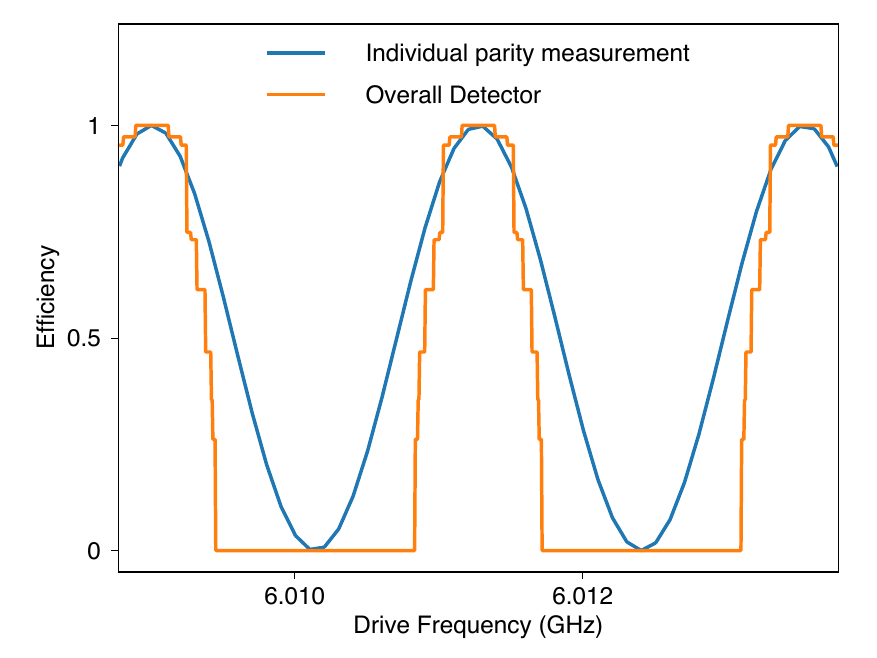}
    }
    \caption{\textbf{Parity measurement and detector efficiency.} The efficiency of an individual parity measurement (blue) is sinusoidal in the frequency of the hidden photon induced drive $\omega_{\gamma '}$. The detector is a series of 30 repeated parity measurements and operates with an efficiency shown in orange.}
    \label{fig:det_eff}
    
\end{figure}

The photon dependent shift of the qubit transition as a function of the frequency of an external drive is determined in Gambetta et. al. \cite{Gambetta_2006} to be $2\chi + \omega_c - \omega_{\gamma '}$ where $\hbar \omega_{\gamma '} = m_{\gamma '} c^2$. The efficiency of an individual parity measurement for a photon dependent shift that is incommensurate with the nominal shift $2\chi$ is given by $\eta_{\mathrm{parity}} = |\frac{1}{2}(e^{i \pi(2\chi + \omega_c - \omega_{\gamma '})/2\chi }-1)|^2$ (Fig. \ref{fig:det_eff}). The effect of an inefficient parity measurement is modeled as a higher probability of qubit error in the hidden Markov model. The data is then reanalyzed and the efficiency of detection in the presence of the additional error is extracted (Fig. \ref{fig:det_eff}).

We note that for large detunings of the external drive, the shifted qubit transition frequency is out of the band of the $\pi/2$ pulses used in the parity measurement. The pulse shapes are Gaussian with $\sigma = \SI {6}{\nano \second}$. This constrains the maximum addressable dark matter detuning from the cavity.

A hidden photon candidate that could result in more detector counts than background counts is only possible if the population of the odd number states of the cavity state ($P_{\mathrm{odd}}$) induced by the hidden photon is larger than the excluded hidden photon probability ($\bar{n}_{\mathrm{HP}}^{90\%}$). To calculate this $P_{\mathrm{odd}}$ we again follow Gambetta et. al. \cite{Gambetta_2006}.

\begin{equation}
P_{\mathrm{odd}} = \frac{1}{\pi} \sum_{k=0}^{\infty}  \mathrm{Re} \left( \frac{\frac{1}{(2k+1)!} (-A)^{2k+1} e^A}{2(2 \pi/T_2^{q, \mathrm{echo}} + \Gamma_m) + (2k+1) 2\pi/T_1^c} \right)
\label{eqn:P_odd}
\end{equation}

where $A = D \frac{\pi/T_1^c - i\chi - i (\omega_c-\omega_{\gamma '})}{\pi/T_1^c + i\chi + i (\omega_c-\omega_{\gamma '})}$ and $\Gamma_m = D \frac{\pi}{T_1^c}$ with the distinguishability $D = \frac{2(n_- + n_+)\chi^2}{(\pi/T_1^c)^2 + \chi^2 + (\omega_c-\omega_{\gamma '})^2}$. $n_-$ and $n_+$ are related to the drive strength ($n_{\mathrm{drive}}$) in units of photons: $n_{\pm} = \frac{n_{\mathrm{drive}} (\pi/T_1^c)^2 }{(\pi/T_1^c)^2 + (\omega_c-\omega_{\gamma '} \pm \chi)^2}$. At a given hidden photon mass, we calculate all $n_{\mathrm{drive}}$ such that $P_{\mathrm{odd}} \geq \bar{n}_{\mathrm{HP}}^{90\%}$.

We note that for external drives with large amplitudes, the shifted qubit transition frequency will be out of the band of the $\sigma = \SI{6}{\nano \second}$ Gaussian $\pi/2$ pulses used in the parity measurement. This constrains the maximum addressable dark matter induced photon occupation.

\begin{figure}[hbt!]
    \centerline{
    \includegraphics[width=\columnwidth]{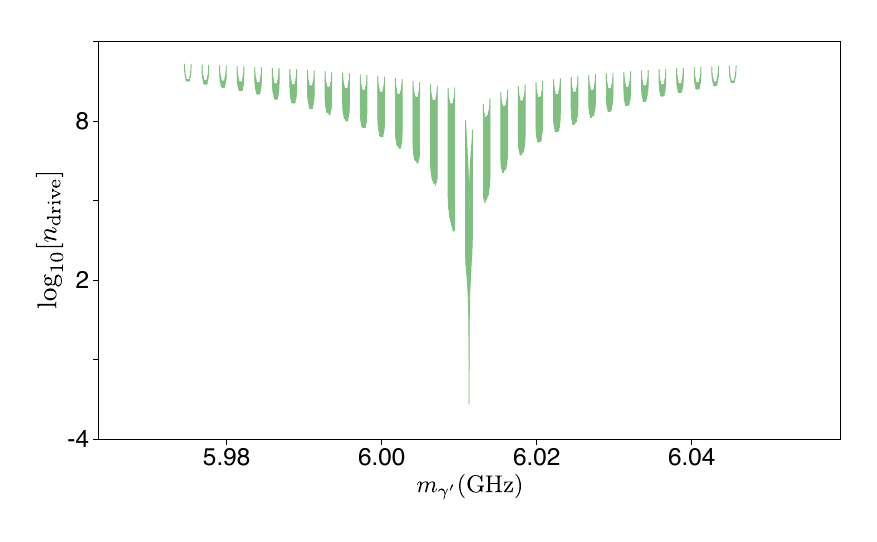}
    }
    \caption{\textbf{Excluded $n_{\mathrm{drive}}$ as a function of $\omega_{\gamma '}$.} The shaded region indicates $n_{\mathrm{drive}}$ induced by the hidden photon that result in $P_{\mathrm{odd}} \geq \bar{n}_{\mathrm{HP}}^{90\%}$ that are detectable and are therefore excluded as possible candidates.}
    \label{fig:n_d_min}
    
\end{figure}

By combining the detector efficiency with the $n_{\mathrm{drive}}$ such that $P_{\mathrm{odd}} \geq \bar{n}_{\mathrm{HP}}^{90\%}$, we determine all $n_{\mathrm{drive}}$ to which the repeated parity measurements are sensitive enough to detect and exclude (Fig. \ref{fig:n_d_min}). Using Eqn. \ref{eqn:N_HP} we convert the excluded $n_{\mathrm{drive}}$ to a region of excluded hidden photon mixing angle ($\epsilon$).

The above calculations assume an infinitely narrow dark matter line. To obtain the excluded region of the hidden photon kinetic mixing angle, we must account for the lineshape of the dark matter \cite{Foster_2018}. We convolve the dark matter lineshape, characterized by $Q_{\mathrm{DM}} \sim 10^6$, with the region shown in Fig. \ref{fig:n_d_min} to obtain the excluded $\epsilon$ shown in the main text.

We note that the storage cavity contains an infinite set of discrete resonances each with a unique coupling to the dark matter. We focus only on the lowest order mode that is specifically designed to couple to the qubit. In principle, the interactions between any modes and the dark matter could result in additional sensitivity to the hidden photon. This would require the mode of interest to have a sufficiently large geometric form factor as well as a resolvable photon number dependent qubit shift. Future dark matter searches could employ structures with multiple resonances to enable multiple simultaneous searches \cite{Chakram_2020_b}.

\section{Future dark matter search}
In order to implement a full scale axion search, the photon counting device must be coupled to a microwave cavity bathed in a magnetic field that accumulates the axion deposited signal. To extract the signal, a nonlinear element such as a Josephson parametric converter can be used to transfer the signal photon from the accumulation cavity to the storage cavity \cite{Axline_2018, Leung_2018}. When the accumulation cavity frequency is tuned to search for a different axion mass, the converter can be pumped at appropriate frequency to enable photon transfer. The storage cavity and qubit can remain fixed in frequency, which leaves the photon detection protocol unchanged at each tuning. Although novel cavity techniques to achieve high Q in the presence of magnetic fields have been demonstrated \cite{Alesini_2020}, in the most pessimistic scenario the accumulation cavity will be made of copper and limited to a $Q \sim 10^4$ at $\SI{10} {\giga \hertz}$ due to the anomalous skin effect. This sets the accumulation time to $\sim \SI{1} {\micro \second}$. To minimize the dead time of the experiment, the time required to measure the storage cavity should ideally be matched to that of a copper accumulation cavity lifetime. Reaching the required detector error probability in this limited time will be challenging. In this work, each parity measurement requires $\SI{10} {\micro \second}$ because of the large readout signal necessary to overcome the HEMT amplifier noise. We perform 30 repeated measurements in order to reduce the probability of detector errors to a level below the expected signal photon probability for dark matter ($\bar{n}_{\mathrm{axion}} \sim 10^{-8}$). Readout of a superconducting qubit with $>99\%$ fidelity in $\SI{100} {\nano \second}$ has been achieved by using quantum limited parametric amplifiers \cite{Walter_2017} and appropriate pulse shaping \cite{McClure_2016}. Both techniques can be applied to this protocol to significantly increase the measurement rate and readout fidelity.

For a hidden photon search, a magnetic field is not required. As demonstrated in this work, the accumulation and storage cavity can be the same device. When the cavity is tuned to search through the parameter space, as long as a sufficiently large dispersive shift to the accumulation/storage is maintained and the qubit is still far detuned, the fundamental QND interaction between the qubit and photon is maintained. Additionally, by using extremely high Q cavities ($Q \gg Q_{\mathrm{DM}}$) to sample the dark matter energy distribution once or twice per dark matter linewidth, only $Q_{\mathrm{DM}} \sim 10^6$ cavity frequency tunings are required to test each mass hypothesis in a frequency octave.



\bibliography{references}





\end{document}